\input cp-aa.tex
\def\teff{T_{\rm eff}}
\def\mast{M_\ast}

\def\msol{M_\odot}
\def\lsol{L_\odot}

\def\mast{M_\ast}
\def\llsol{L/L_\odot}

\def\g{{\it g}}
\def\p{{\it p}}

\def\bv{Brunt-V\"ais\"al\"a }
\def\signorm{\sqrt{3 G M_\ast / R_\ast^3}}
\def\sigr{\sigma_{\rm R}}
\def\sigi{\sigma_{\rm I}}
\def\kapt{\kappa_{\rm T}}

\def\kaprh{\kappa_\rho}

\def\ekin{e_{\rm kin}}

\font\typewr=cmtt10
%
\MAINTITLE{A fresh look into pulsating PG1159 stars
          }
%
%
\AUTHOR {Alfred Gautschy
        }
\INSTITUTE{Astronomisches Institut der Universit\"at Basel,
           Venusstr. 7, CH-4102 Binningen, Switzerland
          }
%
%
\ABSTRACT {
           To improve the understanding of the pulsational
           instabilities of some of the PG1159-type pre-white dwarfs
           we performed detailed stability analyses on helium star
           evolutionary models. With the canonical chemical
           compositions, pulsational instabilities were encountered
           for all stellar masses considered (0.57, 0.63, 0.7
           $\msol$).  The blue boundary agrees satisfactorily with
           observations.  The red edge, on the other hand, is found to
           extend lower temperatures than presently admitted by
           observers.  The dependence of the pulsational instabilities
           on helium or even hydrogen abundance in the stellar
           envelopes is less stringent than emphasized hitherto. In
           particular, we found pulsationally unstable \g~modes for
           model stellar envelopes which are representative for
           HS2324+3944, a newly discovered PG1159-star with a hydrogen
           abundance X $\approx 0.2$ and being pulsating.  
          }
\THESAURUS{ 08(15.1, 23.1, 09.3)
          }
\maketitle %
\titlea {Introduction}
The hottest known stars are in their post-AGB phase shortly before
settling onto the cooling sequence of white dwarfs. Corresponding
stellar evolutionary tracks show a pronounced `knee' joining the
evolution at almost constant luminosity with one at almost constant
radius (cf. Fig.~2).  Spectroscopic observations show that some of the
objects in that domain of the Hertzsprung-Russell (HR) diagram have
high abundances of helium, carbon and oxygen. Characteristic
properties of their spectra are the absence of hydrogen but strong
helium, carbon, and oxygen features (see e.g. Dreizler et
al. 1995). Such stars are named after the prototype PG1159-035. About
50~\% of the 32 known PG1159 stars are surrounded by planetary nebulae
and can be considered as direct descendents of recent leavers of the
AGB. Those without indications of associated nebulosities cannot be
attributed unambiguously an evolutionary state.

Some of the PG1159 stars are photometrically variable with periods
ranging from 10 to about 30 minutes (also termed GW Vir or DOV stars,
we will use the first name in the following). Nine variable stars of
this class are known by now. The high-luminosity members of the class
show typically about four times longer periods than those being close
to the white dwarfs' cooling tracks.  The longer-period PG1159 stars are
typically variable central stars of planetary nebulae and are referred
to as PNNVs or variable CSPNe in the literature. Since there are also
cooler variable central stars it might be appropriate to stress that
we consider here the {\it hot} PNNVs with $\teff$ around $10^5$~K. In
this paper we will not further distinguish between the high- and
low-luminosity branch of the evolution track. Hence, in the
discussions of the mode physics it is of no further importance if
a GW Vir variable is surrounded by a planetary nebula or not. For
aspects of evolution and/or population statistics of the instability
region the very presence of a planetary nebula might of course help to
discern between different scenarios.

The GW Vir variability is attributed to the overstability of
low-degree, high-oder gravity (\g) modes. The observed frequency
spectra are very rich~--~i.e. many modes are excited
simultaneously~--~and they are used for seismological analyses with
considerable success (e.g. Kawaler \& Bradley 1994). Based on the data
of a WET observing campaign, the power spectrum of the variability of
PG1159-035 itself, for example, allowed the resolution of 125
individual frequency peaks (Winget et al. 1991).  For seismological
studies, excitation physics of the pulsation modes is irrelevant. The
results from adiabatic oscillation calculations~--~neutrally stable
oscillations modes~--~serve to identify the observed peaks in the
power spectra.  Irregularities in the distribution of the oscillation
frequencies provide a tool to probe aspects of the internal structure
of the oscillators.  The main objective of seismological studies is
to constrain the stratification of the stellar envelopes and of course
the accurate measurement of global stellar parameters which are not
directly accessible otherwise.

The physical origin of the pulsational instability of GW~Vir-type was
addressed in the middle of the eighties by Starrfield et al. (1984,
1985). They identified partial K-shell ionization of carbon and/or
oxygen as the destabilizing agent. Due to the rather weak opacity bump
produced by this ionization processes, even minor contamination of the
stellar matter by elements other than C and O are considered as
effectively destroying pulsational instabilities.  In particular, H
and He were banished from the excitation region which were assumed to
be made up of mixtures of C and O only in the studies of Starrfield et
al.  (1984, 1985).  A marginal instability region, shortly before the
knee, was obtained the a mass fraction of 0.1 of He in Stanghellini et
al. (1991). A higher helium abundance could not account for
{\it any} hot variables of GW Vir type.

\begfig 0.5 cm
\figure {1}{The observed distribution of some of the PG1159-035
            stars on the $\teff$~--~$\log g$ plane. Solid dots show
            the spectroscopic calibration of photometrically stable
            stars. The white circles indicate the GW Vir
            variables and the hot, variable central stars of
            planetary nebulae (hot CSPNe). The solid lines derive from
            our stellar evolution calculations of helium stars which
            were performed for this study. The grey area covers the
            observationally defined instability domain of PG1159
            stars.  }
\endfig
 
Since the GW Vir variables are few a reliable observational
determination of the borders of the instability region is difficult
and probably still rather unreliable. An attempt was presented
recently by Werner et al. (1995). The grey area in Fig.~1
approximating the instability region is based on Werner \& Rauch's
data; theirs, however continues to lower $\log g$ values which are not
of interest for us since the most massive star we consider has $0.7
\msol$.

Recently, Saio (1996) reviewed the pulsation properties of
hydrogen-deficient stars. Therein he also dealt with GW Vir variables
and hot, variable CSPNe. Using the new generation opacity data (the
OPAL brand in his case) he encountered pulsational instabilities for
chemical compositions which were not found when using the old opacity
tables.  Saio's computations suggest that the previously stringent
chemical composition constraints for pulsational instabilities need to
be relaxed. He presented a period~--~$\teff$ diagram which shows
convincing overlap with observational data.

A major unresolved problem is posed by the observation that a PG1159
star located within the instability region, lets say on the $\log
\teff$~--~$\log g$ plane, does not compulsively imply its
variability. Based on the presently known PG1159 stars, only about
half of the population encountered in the instability domain
pulsates. The dilemma intensifies as we know of spectroscopic twins of
which one is variable and the other stable (Werner et al. 1991).

This paper presents a nonradial, nonadiabatic survey of low-degree
g-modes in GW Vir-like stars using the new OPAL opacity data for the
radiative opacities.  Partially, it can be considered as supporting
the recent results of Saio (1996).  The modeling methods are described
in Sections~2 and 3.  We emphasize aspects of stellar and mode physics
and to some degree the influence of the chemical composition mainly on
a qualitative level rather than trying to fit particular observations
in Sects.~4 and 6. One exception is the star HS2324+3944 which was
recently discovered to be a PG1159 star which shows measurable amounts
hydrogen in its spectrum {\it and} additionally it is a pulsating
variable. Section~5 deals with our modeling of this object.

\titlea {Construction of the PG1159 models}
The stellar background models on which the stability analyses were
performed relied on results from stellar evolution computations. The
nonradial oscillation analyses were hence performed on complete
stellar models. The finite-difference scheme used solves the
structure/evolution equations from the center to the photosphere of
the models. The physical processes accounted for in the code are the
same as described in the paper of Gautschy et al. (1995). In
particular, we point out that He, C, {\it and\/} O ionization are
accounted for in the equation of state (EoS). All regions of the
stellar interior are treated with the identical formulation of the EOS
to avoid discontinuities in thermodynamical quantities at switching
points of the formulations. Convection was dealt with in the standard
MLT approach with a mixing length of 1.5 pressure scale height. For
the very hot stars in which we are interested in here the particular
formulation of convective energy transport should not be very crucial.
Opacity data were taken from the freely accessible CO enhanced OPAL
repository (in its 1992 version) at Livermore ($\kappa_{\rm table}$).
At very high temperatures, above $10^8$~K, the tables were frequently
not broad enough in the $R$ variable to accommodate the stellar
interiors. In such high-temperature domains we extended the tables
with approximations by analytical formulae as described in Gautschy et
al. (1996)($\kappa_{\rm analyt.}$). To avoid jumps in the Rosseland
mean and its partial derivatives we defined an ad-hoc overlap region
in temperature over which we switch linearly, according to
$\theta \cdot \kappa_{\rm table} + (1 - \theta) 
        \cdot \kappa_{\rm analyt.}
$, 
between the two approaches.  The transition width spanned usually
about $10^6$~K. To make best use of the OPAL data we also implemented
their interpolation code ({\typewr cotrin}) interpolating within a
single opacity table as well as between different chemical
compositions for a fixed amount of hydrogen.  We included nuclear
burning of He to C, O, and Ne. Energy losses by neutrinos were
included with the fits of Munakata et al. (1985). Elemental diffusion
in the strong gravitational field building up after passing the knee
was not included in the calculations.

\begfig 0.5 cm 
\figure {2}{Loci of the initially chemically homogeneous helium
            stars with 0.57, 0.63, and $0.70 \msol$ on the HR
            diagram. Stability analyses were performed on models
            having passed their maximum luminosity. The
            internal structure of such models resembles that of
            models coming directly from the AGB and having lost
            their H-rich envelope by some mechanism. The overlaid
            heavy line indicates regimes where pulsationally
            overstable $\ell = 1$ modes were found.
           }
\endfig

Our model series started as homogeneous helium main-sequence stars
with a composition compatible with observed surface abundances in
PG1159 stars. Except if noted otherwise, we used X = 0, Y = 0.38, C =
0.4, O = 0.2. Figure~2 shows the evolutionary tracks of the three
stellar masses ($0.57, 0.63$ and $0.7 \msol$) which were evolved for
the purposes of this study. The chemically homogeneous helium zero-age
main-sequence models were followed as far as possible through the core
and shell helium burning. During the He-shell burning episode they
approach the post-AGB tracks of originally H-rich `canonical' stellar
models. By the time the effective temperature has risen to about
$10^5$~K the position of the He stars on the HR diagram and their
interior structure is close to results from canonical stellar
evolution. Therefore, we are confident that our He star models provide
access to rather realistic structures for PG1159 stars to perform
stability analyses on. Figure 1 indeed confirms that our evolutionary
tracks (for phases after having passed the maximum luminosity) compare
well with the tracks from standard stellar evolution when they cross
the GW Vir instability domain (cf. Dreizler et al. 1995).  The $0.6
\msol$ locus occurring on the $\log g$~--~$\teff$ plane of Fig.~1 is
the result of early test computations and was not used for further
nonradial oscillation analyses. Figure~2 shows the evolutionary tracks
of the three stellar masses ($0.57, 0.63$ and $0.7 \msol$) which were
evolved for the purposes of this study.  The more massive two
sequences run out of the opacity tables at rather low temperatures
already (a few times 1$0^7$~K) shortly after the knee so that the
computations were stopped at that point. The $0.57 \msol$ could be
followed very far down the cooling track. We terminated its evolution
calculations after it reached $\log \teff = 4.7$ at a luminosity of
$1.7 \lsol$.

\begfig 0.5 cm
\figure {3}{Example of the chemical stratification of a $0.7 \msol$
            model passing through the luminosity maximum shown
            in Fig.~1.
           }
\endfig

A representative spatial run of the chemical stratification of the
PG1159 star models is displayed in Fig.~3. It corresponds to a model
from the $0.7 \msol$ series around the luminosity maximum. The burning
shell starts at about $0.8 \mast$. Within about a tenth of the stellar
mass helium is burned away. As evolution proceeds the burning shell
propagates outward. Except for a translation, the functional form of
the abundance profile does not change significantly.  The same applies
when considering different stellar masses.

\titlea {Nonradial nonadiabatic oscillation treatment} 
The nonradial, nonadiabatic stellar oscillation code was the same as
described in the paper of Gautschy et al. (1996). The nuclear terms
were included here for the sake of completeness.  As helium burning
only was considered, equilibrium treatment of the $\epsilon$-mechanism
was sufficient. The boundary conditions in the center are dictated by
the regularity of the perturbations of the physical quantities.  At
the surface we assume that the Stefan-Boltzmann law for radiation
obtains and full reflection of the waves as the mechanical boundary
condition.  Convection was always treated as being frozen-in, ie. the
perturbation of the convective flux was assumed to be negligible. In
most of the hot models the particular treatment of convection might be
irrelevant due to its marginal contribution to the energy transport. A
more detailed discussion of this point is postponed to Sect.~7 .  The
eigenvalues $\sigma = (\sigr, \sigi)$, where $\sigr$ stands for the
oscillation frequency and $\sigi$ for the excitation/damping rate, are
always expressed in units of $\signorm$.

\titlea {Results from stability analyses} 
In this section we present the results of nonradial, nonadiabatic
pulsation computations performed on $0.7, 0.63$, and $0.57 \msol$
evolutionary sequences. Each of the model series will be dealt with
separately. We start by discussing the obtained {\it modal diagrams\/}
(plot of period vs. some control parameter such as $\log g$ or damping
rate of a particular mode) and consider consequences
therefrom. Mostly, the mode computations were restricted to $\ell = 1$
\g~modes.

We start with the most massive star sequence~--~the $0.7 \msol$ models.
Figure~4 shows the variation of the periods of some of the $\ell = 1$
\g~modes as evolution proceeds. The star's evolution is parameterized 
by its surface gravity which monotonically increases advancing
evolutionary stage in the phases covered by us.
\begfig 0.5 cm
\figure {4}{
            Modal diagram of the lower-order, $\ell = 1$ \g~modes of
            the last part of the $0.7 \msol$ sequence. Top panel:
            Light lines trace the variation of the oscillation periods
            of particular overtones as the surface gravity changes
	    during the evolution. Unstable regimes are marked 
            by thick lines. The short period modes are separated by
            $\Delta k = 1$ up to the mode pointed to with the
            appropriate error. The long-period modes are, as indicated
	    with the other arrows, separated by $\Delta k = 10$. An
            exception is the mode with the asterisk. Its neighboring
            modes are separated by $\Delta k = 5$.
            Bottom panel: the variation of
            the imaginary parts of the modes. Due to the large range
            spanned by the imaginary parts the scaling is
            logarithmically for values much larger than $10^{-8}$.
           }
\endfig

The upper panel of Fig.~4 shows the variation of the periods of some 
\g~modes as the star evolves around the knee. A particular radial
order is traced by a thin line.  Since we were dealing with high-order
modes a small fraction only of the radial orders (denoted by $k$) in
the period domain of interest for GW Vir variables was followed. In
this respect, the upper panel can be divided again. In the low-period
domain~--~between about 50 and 200~s~--~we computed eigenvalues for
every radial order.  Above about 300~s~--~in the long-period
regime~--~only every 10th radial order (region between the arrows
originating from $\Delta k = 10$ in Fig.~4) was followed. The only
exception with $\Delta k = 5$ to both sides is marked with an
asterisk. For easier identification of the instability domains, the
pulsationally unstable mode branches are traced with heavy lines.
From the lower panel in Fig.~4 we see this to correspond to negative
$\sigi$ values in our sign convention of the eigenvalues. For better
graphical presentation we appropriately scaled the large numerical
range of the imaginary parts. Moduli of $\sigi$ exceeding $10^{-8}$
scale logarithmically in this choice of the ordinate. Smaller values
behave essentially linear. The sign function allows to discern between
stable and overstable oscillation modes.
 
The short-period domain in Fig.~4 shows sequences of avoided crossings
occuring in low-order modes of low surface-gravity models. The avoided
crossings appear in period as well as in the imaginary parts of the
eigenfrequencies. A local analysis of the first model of the series
shows that eigenmodes with periods below about 220~s have dual
character. Deep in the interior they propagate as \g~modes and as
\p~modes in the envelope.  The less steep portions of the eigenmode
branches which cause the avoided crossings reflect the different
reaction on the stellar evolution of the \p-mode propagation speed in
the envelopes. Away from avoided crossings, the less steep branches
correspond to modes which are dominated by the \p-mode cavity. In
accordance with a strong coupling with the envelope of the
short-period modes early in our evolutionary phases they have,
compared with core-dominated \g~modes, high damping rates which are
induced by considerable radiative damping in the envelopes. Figure~4
shows imaginary parts as high as $10^{-5}$ only. The largest $\sigi$
values rise to about $10^{-3}$.

For a short phase of the evolution, one of the short-period modes
turns weakly pulsationally unstable. The period ranges from 145 to
152~s. The excitation rate is very weak as can be seen in the lower
panel of Fig.~4. We ensured its reality by re-iterating this mode
several times with different accuracy bounds for the numerical scheme
and with different start-guesses for the eigenvalue. Repeated
convergence to this weak instability proved that we can confirm the
pulsational instabilities at short periods which were also seen by
Saio (1996, his Fig.~5). The instability is induced by the He-burning
shell through the action of the $\epsilon$-mechanism. Only a very
narrow period range and a very short evolutionary phase are prone to
such instabilities. In the $0.7 \msol$ star models this leads to one
radial order only which can be weakly destabilized at $\ell = 1$. The
radial displacement eigenfunction of the unstable mode has 5 nodes
within the \g-mode cavity and none in the acoustic propagation region
which it hardly touches anymore.  The maximum growth rate of the
nuclearly destabilized \g~mode amounts to an e-folding time of $1.5
\cdot 10^3$~y. The star's evolution through this instability phase
takes about 10 times longer. This might be marginally sufficient for
an $\epsilon$-destabilized mode to build up.  In comparison with the
long-period g\-mode instabilities, the instabilities effected by the
$\epsilon$-mechanism appear to be hardly of relevance.

In the long-period domain, where modes modes have always pure \g-mode
character, pulsational instabilities occur for periods longer than
880~s (cf. Fig.~4). The higher the radial order of the modes the lower
the surface gravity for which instability shows up.  The locus of the
mode marked with `$\ast$' has $k = 45$. It was inserted in the usual
$\Delta k = 10$ sequence to more accurately determine the low-period
border of the instability region. Figure~4 shows a small fraction only
of all the modes that are be pulsationally unstable.  We restricted
the computation of overtones as high as necessary to find stable modes
again to the first stellar model of the particular model sequence. The
longest-period mode which was unstable in the $0.7 \msol$ sequence
lies at about 7447~s. This is a truly remarkable range of
simultaneously unstable radial orders.  Since the Riccati method does
not involve the simultaneous computation of the eigenfunctions when
relaxing eigenfrequencies we estimate the range of radial orders
covered by the asymptotic \g-mode period separation formula. For the
model at $\log g = 5.94, \log \teff = 5.045$, and $\ell = 1$ we deduce
about 200 modes to be simultaneously unstable. The return to stability
of the very-high overtones appears rather abruptly. Over most of the
unstable period range, the imaginary parts remain at $\sigi = {\cal
O}(10^{-6})$.

\begfig 0.5 cm
\figure {5}{
            Same as Fig.~4 but for the $0.63 \msol$ series. The
            outlined numbers are locations where different spherical
            degrees were computed; they are discussed in Sect.~6.
            The horizontal line at $\log g = 6.745$ indicates the
            position where period separations and the variation of 
            the imaginary parts of eigensolutions of subsequent radial
            orders.
           }
\endfig

The models of the $0.63 \msol$ evolutionary track were processed
similarly as those of the $0.57 \msol$ sequence. The resulting $\ell =
1$ modal diagram is shown in Fig.~5. Only the modes relevant for the
eventually uncovered instability region are shown. The modal analysis
of the $0.63 \msol$ models is the most dense one~--~in terms of
covered radial orders~--~of this project. The radial-order spacing
between the displayed lines in the top panel of Fig.~5 are labeled.

As in the $0.70 \msol$ case, the instability region has an expressed
finger-like shape pointing from low-$\log g$ and long periods to
high-$\log g$ and short periods. The high-$\log g$ edge of the
instability region occurs at about $\log g = 6.85$. The hottest point
of the $0.63 \msol$ evolutionary track is encountered at $\log g =
7.0$. Hence, the instability domain terminates before $0.63 \msol$
stars start to dim their luminosities. Our computations showed no
return to pulsational instability of the evolutionary models during
their `downward' evolution anymore (as indicated in Fig.~1). 

Again, a large number of radial overtones remained overstable also in
the earliest model of the sequence and extending again to very long
periods.  Therefore we stopped the tracking of particular radial
orders around 5000~s. The very smooth and stretched ridge in
$\sigi$ is seen in Fig.~7's lower panel. In contrast to classical
pulsators such as Cepheids or RR Lyrae stars, a large number of
radial overtones appear to be simultaneously favored for excitation. The
highest-order $\ell = 1$ mode which was found overstable at $\log
g = 5.83$ has a period of about 9500~s.

The locus of the $k = 91$ dipole mode is marked in Fig.~5. This serves
to count the radial orders of the other modes plotted and to emphasize
the high order of excited modes. The number of nodes was counted in
the displacement eigenfunction. This particular mode was chosen
because it belongs to those modes with maximal instability in the $0.63
\msol$ sequence. We return to in Sect.~6.

\begfig 0.5 cm
\figure {6}{
            Same as Fig.~4 but for the $0.57 \msol$ series. The
            dashed horizontal line marks the hot-most point
            of the evolutionary knee. For higher surface gravities
            the star approaches the white dwarf cooling tracks. 
           }
\endfig

The lowest-mass evolutionary sequence considered here had $0.57
\msol$. The results from the nonadiabatic eigenanalyses for $\ell = 1$
are shown in Fig.~6. In contrast to the before-mentioned sequences,
here the instability domain extends beyond the knee into the cooling
phase.  Therefore we indicated the to highest effective temperature
reached with a horizontal line at the appropriate $\log g$ value. For
economic reasons we restricted ourselves to a rather coarse coverage
of the period domain. The low period range ($\Pi \la 800 $~s) was
scanned with $\Delta k = 5$ until we reached that radial order which
did not turn overstable during the evolutionary phase depicted in
Figs.~1 or 2. At long periods we stopped the tracking of modes after
the first model of the $0.57 \msol$ sequence turned pulsationally
unstable.  For the first model we searched the longest overstable mode
which was found at about 5500~s.

The long-period `instability finger' is also contained in Fig.~5.  It
merges, however, with a short-period instability domain which extends
to low luminosities and hence late evolutionary stages. Periods as
short as 300~s become overstable. The $\ell = 1$ \g~modes of the last
model of the $0.57 \msol$ evolutionary sequence were all pulsationally
stable again. Hence, the red edge on the low-luminosity branch lies at
$\log \teff = 4.45$ and $\log \llsol = -0.1$. In our computations
there is one radial order (the third longest) which which becomes
pulsationally overstable at high luminosities and again~--~after a
phase of stability~--~at low luminosities.

The lower panel of Fig.~6 shows the variation of the imaginary parts
of the eigenfrequencies. The high-luminosity part is smooth and looks
very similar to the results of the more massive star sequences. The
low-luminosity imaginary parts are, despite their comparable modulus,
less smooth.  This is not numerically caused. At low luminosities the
transition from the He-rich envelope to the pure CO core is rather
steep and as close to the surface as it can ever come. Therefore, the
\g~modes are partially reflected at the composition transition such
that trapping of modes could be expected. Such trapping of modes also
shows up in the imaginary parts of the eigenfrequencies (cf. Gautschy
et al. 1996) leading to the `bumpy' appearance. We finally notice that
the width of the instability with respect to radial orders is much
narrower at low luminosities than on the high-luminosity branch. The
envelope covering the short-period $\sigi$ peaks in the lower panel of
Fig.~6 crosses zero at a very steep angle at about 1000~s. This is
much different from the behavior of the envelope of the $\sigi$ curves
in larger-mass sequences for which the long periods are unstable only
on the high-luminosity branch of the evolutionary track. 

\titlea {The peculiar case of HS2324+3944}
Dreizler et al. (1996) reported the discovery and spectroscopic
analysis of the peculiar PG1159 star HS2324+3944. In contrast to the
other members of this class it shows an atmospheric ratio (by number)
of He to H of 0.5. Furthermore, HS2324 shows an unusual low oxygen
abundance. In mass fractions, we adopt X=0.2, Y=0.41, C=0.37, O=0.01
(Werner, private communication).  Recently, Silvotti (1995) provided
first observational results indicating that HS2324 is a GW~Vir-type
variable with a dominant period of about 2140 s.  According to previous
nonradial stability analyses (e.g. Starrfield et al. 1984) any trace
of hydrogen was considered poisoning pulsational instabilities.

The spectroscopic calibration places HS2324 at about $\log g = 6.2$
and $\log \teff = 5.11$ (Dreizler et al. 1996).  Based on the loci of
our helium-star models in the $\log g$~--~$\log \teff$ plane, we
assumed a stellar mass of $0.63 \msol$, this is slightly larger than
what Dreizler et al.  proposed.  We constructed stellar envelope
models with homogeneous chemical composition adopting abundances as
mentioned in the last paragraph. The integration starting at the
stellar surfaces extended to the onset of significant hydrogen
burning at about $3 \times 10^7$~K. 

\begfig 0.5 cm
\figure {7}{
            Comparison of pulsationally interesting physical 
            quantities between $0.63 \msol$ He star models (pg63 seq.)
            and envelope models for HS2324. The HS2324 models differ
            in their chemical composition, in particular the 
            under-abundance in O compared with the evolutionary
            sequences.
            Top panel: logarithmic opacity derivative with respect
            to temperature at constant density. Bottom panel:
            spatial run of the logarithm of the opacity. The 
            depth in the stellar envelopes is parameterized by the
            temperature.
           }
\endfig

The bottom panel of Fig.~7 clearly shows the decrease of the opacity
in the envelope models with HS2324 composition when compared with the
evolutionary helium star models. The opacity bump around $\log T =
6.25$ appears to become narrower. Hence, pulsational driving is not
expected to be significantly weakened by neither the existence of
hydrogen nor the suppressed oxygen abundance. The $\kapt$ slope in
both stellar models (HS2324 envelope and evolutionary He-star) are
comparable (cf. top panel of Fig.~7). Actually, the slope in the
HS2324 model appears to be marginally steeper than in the full He-star
model.

\titleb {The pulsation properties of HS2324}
As for the full helium-star models, we performed eigenanalyses on the
envelope models for HS2324. The inner boundary of the envelope was
treated as a reflective wall. This influences the period spacing
between adjacent orders of the computed \g~modes only. In the cases
considered the spacing was about five times larger than for full $0.63
\msol$ models with comparable effective temperature. The magnitude
of the work integrals are not expected to be much different. Most of
driving and damping occurs below $\log T = 6.5$. In particular the
decisive contributions to stability or instability come from there. In
HS2324 we miss some dissipation from the deep interior. The \bv
frequency achieves its maximum close to the center so that the
eigenfunctions have there their most rapid spatial oscillations.  From
the full models we derive that only a few \% of the total work is done
in the deep interior. Our imaginary parts are hence expected to be
marginally too large. The qualitative behavior is, however, believed
to be correct.

\begfig 0.5 cm
\figure {8}{
             Appropriately scaled imaginary parts of eigenfrequencies
             as a function of period of $\ell = 1$ \g~modes.
             The two curves show the results of two different choices 
             of the stellar models' $\teff$ both at $\log \llsol = 3.16$.
           }
\endfig

Figure~8 displays the imaginary parts of low-order $\ell = 1$
\g~modes for two different envelope models. Obviously it is 
possible~--~even with the considerable hydrogen admixture of 20~\% in
mass~--~to obtain pulsational instabilities for HS2324-like stellar
envelopes. Figure~8 shows that the destabilized period range depends
markedly on the choice of effective temperature. Based on the results
for $\ell = 1$ and the observed period of the variability we tend to
favor a temperature closer to $\log \teff = 5.07$ which is
lower~--~but still within the limits of uncertainty~--~than suggested
by Dreizler et al. (1996).

As Fig.~8 clearly shows, the period domain over which dipole modes
become pulsationally unstable is much narrower in the HS2324 models
than in the hydrogen-free and oxygen-rich helium-star models.  The
excitation rates are comparable in both cases. Hence, it is not at all
surprising if pulsations develop~--~as they do in GW Vir
variables~--~in HS2324-like objects. The frequency spectrum must be
expected to be poorer, though.

\titlea {Discussion} 
In the following we take a closer look at the properties of the
computed eigensolutions and the physical properties of the driving
behind GW Vir-type instability.

\begfig 0.5 cm
\figure {9} {Work integral and normalized radial displacement
             of the nonadiabatic eigensolution for $k = 91, \ell = 1$
             of the $0.63 \msol$ model at $\log g = 6.43$. This mode
             belongs to the most unstable ones encountered. Blow-ups
             of the central behaviors in displacement and
             contributions to the work integral are shown as small
             inlets.  }
\endfig

The high quality of the nonadiabatic eigensolutions which was achieved
in this study can be seen in Fig.~9. The top panel displays $\rho r^2
(\vert \xi_r \vert + \ell (\ell + 1) \vert \xi_h \vert) \propto e_{\rm
kin}$ which corresponds~--~up to a normalization~--~to $\diff E_{\rm
kin} / \diff r$ where $E_{\rm kin}$ stands for the total kinetic
energy of an oscillation mode. The global run of $\ekin$ shows the
large weight of the deep interior for the kinetic energy of the
mode. This behavior is characteristic of all \g~modes in our
models. The expressed central concentration of $\ekin$ leads to large
values of $E_{\rm kin}$ which, in a quasi-adiabatic sense, explain the
small values of $\sigi$.  Even if the positive and negative parts of
the work integral do not nearly cancel. The local depression in the
amplitude of $\ekin$ around $\log P = 21.5$ coincides with the inner
edge of the shell-burning region and the occurring abundance bump in
carbon (which is also visible in Fig.~3). Mode trapping can be
associated with such features. For a recent study of ZZ Ceti stars
touching this aspect see Gautschy et al. (1996). The effect is so
weak, though, that it is not of much relevance for observational
properties of GW Vir stars. Mode trapping will be addressed further
down in more detail.

 
The bottom panel of Fig.~9 displays the total work done by the
particular oscillation mode. The dominant driving and damping
contributions lie between $\log P = 10$ and 14. A more
detailed physical discussion follows. The rapid oscillations of the
eigenmodes deep in the star lead to a small damping contribution of
the total work only. A blow-up of the deep interior shown in the inlet
in the lowest panel quantifies this statement.
 
\begfig 0.5 cm
\figure {10}{Eigensolution components and stellar envelope properties
             of the same model and the same mode as used for Fig.~9.
	     To concentrate on the main driving and damping regions of
	     the oscillations, we restrict the plot to temperatures
	     below  $\log T = 7.5$.}
\endfig

Figure~10 shows selected quantities from nonadiabatic nonradial
analysis of the dipole mode $k = 91$ as well as from the stellar
envelope structure of the underlying model (at $\log g = 6.34$ of the
$0.63 \msol$ sequence). This figure is the basis of a more detailed
physical discussion of mode excitation.

The top panel of Fig.~10 depicts the superficial regions of the radial
component of the displacement eigenfunction. It demonstrates that most
of the spatial oscillations of the eigenfunctions are not
significantly involved in the excitation/damping of the mode. The
differential work $\diff W / \diff r$ is seen to be significant in the
range $5.8 < \log T < 6.8$. Mode-driving is restricted to the range
between 5.8 and 6.4 in $\log T$ which is in accordance with the
opacity bump which reaches a local maximum around $\log T =
6.25$. This feature which is due to the combined effects of L-shell
transitions of Fe and K-shell transitions C, O, and Ne (cf. Seaton et
al. 1994) was found to be slightly enhanced (about 20~\%, Iglesias et
al. 1992) in the new tabulations of stellar opacities (OPAL and
OP). Mainly, however, its shape became more pronounced so that it
stands out even more clear in a plot of $\kappa_{\rm T} =
\partial \log \kappa / \partial \log T \vert \rho$. Figure~7
demonstrates the influence of strongly reducing the oxygen abundance
in the stellar matter. The opacity is globally decreased, the bump
around $\log T = 6.2$ decreases also. The sharpness of its ridge on
the $\log R$~--~$\log T$ plane is, however, hardly influenced.
Therefore, the $\kapt$ run is not very much affected by reducing the O
abundance. From the spatial ionization structure of the stellar
envelope used for Fig.~10 we see that partial K-shell ionization of
carbon might contribute only in the outer parts of the envelope, at
temperatures below $10^6$~K. Mostly oxygen K-shell ionization and
bound-bound transitions in Fe dominate the hotter, more important part
of the work integral.  We admit that the comparison of the EoS and the
opacity data is not fully satisfactory. We were not able to use
the OPAL EoS tables with our high C and O abundances for the
discussion. Hence, we must rely on our simplified approach
which is, hopefully, not too far from reality (cf. Gautschy et al. 1996).

The $\kappa$-effect consists of two terms which contribute
to driving if 
$$
  {\diff \over \diff r} \left ( \kapt + { \kaprh \over 
                                          \left ( \Gamma_3 - 1 \right )
                                        }
                        \right ) > 0.
\eqno(1)
$$
From the panel in Fig.~10 which displays the second term in the
bracket it is obvious that its gradient is small throughout the
driving region and it contributes~--~if at all~--~in the
low-temperature part only. Below $10^6$~K, $\Gamma_3 - 1$ drops
noticeable only when the ionization stages O\,{\sc viii} {\it and\/}
O\,{\sc ix} are already considerably populated. At higher temperatures
$\Gamma_3 - 1$ is completely flat so that the first term of Eq.~1
controls the driving peak of the differential work.

\begfig 0.5 cm
\figure {11}{ Period separation and variation of imaginary parts
              of successive radial orders for the $0.63 \msol$
              model at $\log g = 6.745$ (indicated by the horizontal
              line in Fig.~5) for $\ell = 1$.
           }
\endfig

The period separation and the variation of the imaginary parts as a
function of successive radial orders of dipole modes of a selected
$0.63 \msol$ model are shown in Fig.~11. This particular stellar model
is close to the hot-most point of the corresponding evolutionary
track.  Its position is indicated by a horizontal line in the modal
diagram of Fig.~5.  In the period range between 600 and 1200~s no
obvious sign of mode trapping is visible. In particular, no trapping
shows up in the imaginary parts (much in contrast to the way trappings
appear in ZZ Ceti stars, cf. Gautschy et al. 1996). The
oscillation-period separation ($\Delta \Pi$) as a function of period
shows small, seemingly uncorrelated variations on the level one
second. The reason is either intrinsic or then reflects the achieved
numerical accuracy. The asymptotic formula for low-degree, high-order
modes predicts a $\Delta \Pi \approx 23.7$~s which is in reasonable
agreement with the full solutions. An averaged period-separation curve
might possibly hint at a trapping cycle which would then, however,
considerably exceed 600 s.

\begfig 0.5 cm
\figure {12}{ Reaction of the imaginary parts of eigenvalues
              of selected $0.63 \msol$ models (outlined numbers
              in Fig.~5) on varying the spherical degree $\ell$.
            }
\endfig

Figure~5 contains the numbers one to four in an outlined font.  At
these locations we investigated the dependence of the eigensolutions
on varying the spherical degree $\ell$. Mathematically, $\ell$ can be
treated as a continuous variable which was done to generate
Fig.~12. The imaginary parts resulting from such a variation are
displayed in Fig.~12. Notice that in contrast to other plots of
$\sigi$ values, the ordinate has a linear scale. Except at location 2,
dipole modes appear to be the most unstable ones. Only at the tip of
the instability finger, i.e. at position 2, the marginal instability
of dipole modes strengthens towards higher $\ell$ values and achieves
a maximum at about $\ell = 2.4$.  From the observational point of
view, we expect $\ell = 1$ {\it and} $\ell = 2$ modes to dominate
the oscillation spectra.

\titlea {Conclusions} 

We computed three sequences of helium stellar models (at 0.57, 0.63
and $0.7 \msol$ which passed through $\log g$~--~$\teff$ the domain which
is also populated by GW~Vir variables.  A large number of mostly
dipole oscillation modes were investigated towards their stability
properties with a fully nonadiabatic pulsation code. 

GW Vir-like pulsations are driven, as we know for a while already
(e.g. Starrfield et al. 1984) by the $\kappa$-mechanism of partial
ionization of oxygen and carbon. Our Fig.~10 shows the associated
opacity bump which peaks around $1.7 \cdot 10^6$~K and which is
somewhat enhanced in the new opacity tables (OP: Seaton et al. 1994,
OPAL: Iglesias et al. 1992) compared with old data. It appears that
not only detailed atomic physics of partial K-shell ionization in
carbon, oxygen, and neon are relevant but that spin~--~orbit coupling
of L-shell transitions in iron increases the sharpness of the bump
feature additionally.  The literature was, in our opinion, never very
specific on the particular roles, carbon and oxygen played in the {\it
final\/} picture. According to our EoS, which is not the same as that
used in the opacity computations, it should be mainly partial
ionization of the K-shell of oxygen which influences the driving of
the pulsations. The last ionization level of carbon can only influence
the low temperature flank of the driving regime but it cannot dominate
the driving. This conclusion should of course be tested at some point
in detail with consistent EoS and opacity data.  As seen in the case
of HS2324, carbon and oxygen abundances do not yet fully determine the
problem. For the new variable HS2324, we assumed the oxygen abundance
by mass to be as low as 0.01 and hydrogen to make up 20~\% in mass. We
still found pulsational instabilities.  The domain of overstable modes
was, though, much narrower than in the case of `canonical' GW Vir
compositions. Nevertheless, we conclude that the stringent composition
requirements which were put forth hitherto (e.g. Starrfield et
al. 1985, Stanghellini 1991) seem no longer to obtain with the new
opacity data.  Together with the $\kappa$-mechanism the
$\gamma$-mechanism is frequently brought into play to explain GW~Vir
variability. The $\gamma$-mechanism is connected with the spatial
variation of $\Gamma_3 - 1$ in the second term of eq.~1. In the case
of the GW~Vir models, though, the K-shell ionization region of carbon
and oxygen reduces $\Gamma_3$ only marginally. In our pulsationally
overstable models, maximum driving occurs at depths of the last
partial ionization stage of oxygen for which$\Gamma_3$ remains
essentially constant. The second term of eq.~(1) was shown as the
third panel from below of Fig.~10. From this we conclude that the
$\gamma$-effect is not effective for the action of the
$\kappa$-mechanism in the GW~Vir variables~--~an unusual situation
in the theory of pulsating variable stars.

The instability region for dipole modes is shown with heavy lines in
Fig.~2. When translated onto the $\log \teff$~--~$\log g$ plane, we
find reasonable agreement of the blue edge with observations.  As
already found in older studies (e.g. Stanghellini et al. 1991), the
computed red edge occurs at much lower temperatures than what
observations suggest. The $0.57 \msol$ models for which this applies
in our study show only very weak convection zones close to the
surface. A considerable leak of pulsation energy is therefore not
likely to occur with an improved convection-pulsation coupling in our
models. To introduce more efficient convection we would have to change
the MLT prescription or at least assume a mixing-length parameter
which is larger than ours which was chosen to be $\alpha_{\rm MLT} =
1.5$ pressure scale heights.

Furthermore, the longest periods which we found to be unstable
significantly exceed the longest observed ones (around 2000~s). Lower
stellar masses tended to have longer unstable periods. The longest
period which we encountered has about 9500~s and belongs to the $0.57
\msol$ sequence. We saw that dissipation in the deep interior is only
marginal compared to the contributions in the temperature range $5.8 <
\log T < 6.5$.  (cf. Fig.~10). Therefore, we expect the
low-temperature region and the run of relevant components of the
eigensolution therein to be decisive for the stability property rather
than centermost parts of the stars with their very short spatial
wavelengths of the eigenfunctions.

\begfig 0.5 cm
\figure {13}{ Period~--~effective temperature diagram for the 
              $\ell = 1$ modes of the $0.57 \msol$ model sequence. As
              before, pulsationally unstable modes are shown with
              thick lines. For comparison, the two cool GW Vir stars
              PG2131+066 and PG0122+200 are overlaid with a bar
              spanning the observed period domain and indicating the
              $\teff$ uncertainty in the other direction.  }
\endfig

In Fig.~13 we overlaid the observed oscillation period-range and the
spectroscopic temperature calibrations of the cool GW Vir variables
PG2131+066 and PG0122+200 on the same oscillation data as displayed in
Fig.~6. We are interested in how well they fit into the theoretically
determined unstable period domain of the $0.57 \msol$ sequence. We
refer to the $0.57 \msol$ sequence since this is the only of our
sequences which shows unstable oscillation modes at sufficiently low
effective temperatures. We notice that a significantly smaller period
range is observed than what the computations suggest. O'Brien et
al. (1996) adopted a period spacing of 21.2~s for PG0122+200. This is
smaller than the 24.1~s from our models. Increasing the stellar mass
helps to reduce the discrepancy.  Saio's (1996) computations showed
that the `blue edges' of the \g-mode instability regions along the
white dwarfs' cooling tracks are a function of mass. Increasing the
stellar mass shifts the blue edge to lower effective temperatures and
hence lower luminosities ($\log \teff = 5.08$ for $0.58 \msol$ and
$\log\teff = 5.04$ for $0.60 \msol$). Based on the available data we
cannot yet accurately estimate an upper limit of the mass of
PG0122+200. However, the hotter the star actually is the lower is this
upper mass limit to guarantee consistency with the very existence of
\g-mode pulsations.

For none of our models could we find mode-trapping cycles (cf. the
particular case of the $0.63 \msol$ model) that were of relevance for
observed properties in GW Vir variables. The non-smooth run of $\sigi$
at low luminosities of $0.57 \msol$-sequence models marks trapping at
the composition transition of the extinguishing He shell. Due to the
large depth of this composition transition, the possible trapping
length is much too long to be compatible with observational
evidence. Hence, elemental diffusion is a necessary ingredient to
interpret observed unequal period separations.

We still have to live with the enigma of seeing about as many stable
as pulsating stars in the GW~Vir instability domain. This problem is
even somewhat aggravated by the results of this study since the
particularities of the chemical composition are not necessarily as
stringent as believed before. It is conceivable that the heavy element
abundance in these objects has a stronger effect than believed up to
now. If the iron-abundance is low enough, at least in the driving
region, then driving might be reduced considerably even if carbon and
oxygen abundances are PG1159-like. From the point of view of stellar
evolution theory, mode trapping might prove to be a discriminant to
distinguish between {\it first-passers\/} (on their first passage to
the cooling track) and {\it late shell flashers\/}. For first passers
at high enough luminosity or low enough surface gravity sedimentation
by pressure diffusion might not have been very effective yet. This
leads to a lack of $\mu$-barriers which can partially reflect and trap
waves. Such stars are expected to show very regular period spacings
which agree well with simple asymptotic formulae. On the other hand,
low-luminosity stars or late shell flashers had enough time for He
say, to float on top of carbon and oxygen. Even a shell flash need not
necessarily destroy such a stratification if the star does not loop
too far into the low-$\teff$ domain. In such cases, the period
separations should show cyclic depressions as observed or computed
e.g. in PG1159 (Kawaler \& Bradley 1994).

\acknow{I am indebted to H. Saio for many helpful hints and critical
        comments throughout the project. K. Werner and S. Dreizler
        kindly informed me about HS2324 properties prior to their
        publication.  The Swiss National Science Foundation provided
        financial support through a PROFIL2 fellowship which is
        gratefully acknowledged.  }
        
\begref{References}

\ref Bergeron P., Wesemael F., Lamontage R., 
     Fontaine G., Saffer R.A., Allard N.F., 1995,
     ApJ 449, 258

\ref Dreizler S., Werner K., Heber U. 1995, 
     in White Dwarfs, Lecture Notes in Physics 443, 
     eds. D. Koester and K. Werner, Springer, p.~160

\ref Dreizler S., Werner K., Heber U., Engels D., 1996, 
     A\&A in press 

\ref Gautschy A., Ludwig H.-G., Freytag G., 1996,
     A\&A in press

\ref Iglesias C.A., Rogers F.J., Wilson B.G., 1992, 
     ApJ 397, 717

\ref Kawaler S.D., Bradley P.A. 1994, 
     ApJ 427, 415

\ref Munakata H., Kohyama Y., Itoh N. 1985, 
     ApJ 296, 197; Erratum in ApJ 304, 580

\ref O'Brien M.S., Clemens J.C., Kawaler S.D., Benjamin T., 1996,
     preprint

\ref Silvotti R., 1995, 
     IBVS 4265

\ref Saio H. 1996,
     in Hydrogen Deficient Stars, Proceedings of the Bamberg
     conference 1995, in press 

\ref Seaton M.J., Yu Yan, Mihalas D., Pradhan K. 1994,
     MNRAS 266, 805

\ref Stanghellini L., Cox A.N., Starrfield S. 1991,
     ApJ 383, 766

\ref Starrfield S., Cox A.N., Kidman R.B., Pesnell W.D. 1984, 
     ApJ 281, 800

\ref Starrfield S., Cox A.N., Kidman R.B., Pesnell W.D. 1985, 
     ApJ, L23

\ref Werner K., Rauch T., Dreizler S., Heber U. 1995, 
     in Astrophysical Applications of Stellar Pulsation, IAU
     Coll. 155, eds. R.S. Stobie and P.A. Whitelock, ASP Conf. 
     Ser. Vol.~83, 96

\ref Werner K., Heber U., Hunger K. 1991,
     A\&A 244, 437

\ref Winget D.E., Nather R.E., Clemens J.C., 
     Provencal J., Kleinman S.J., et al. 1991, 
     ApJ 378, 326   

\endref
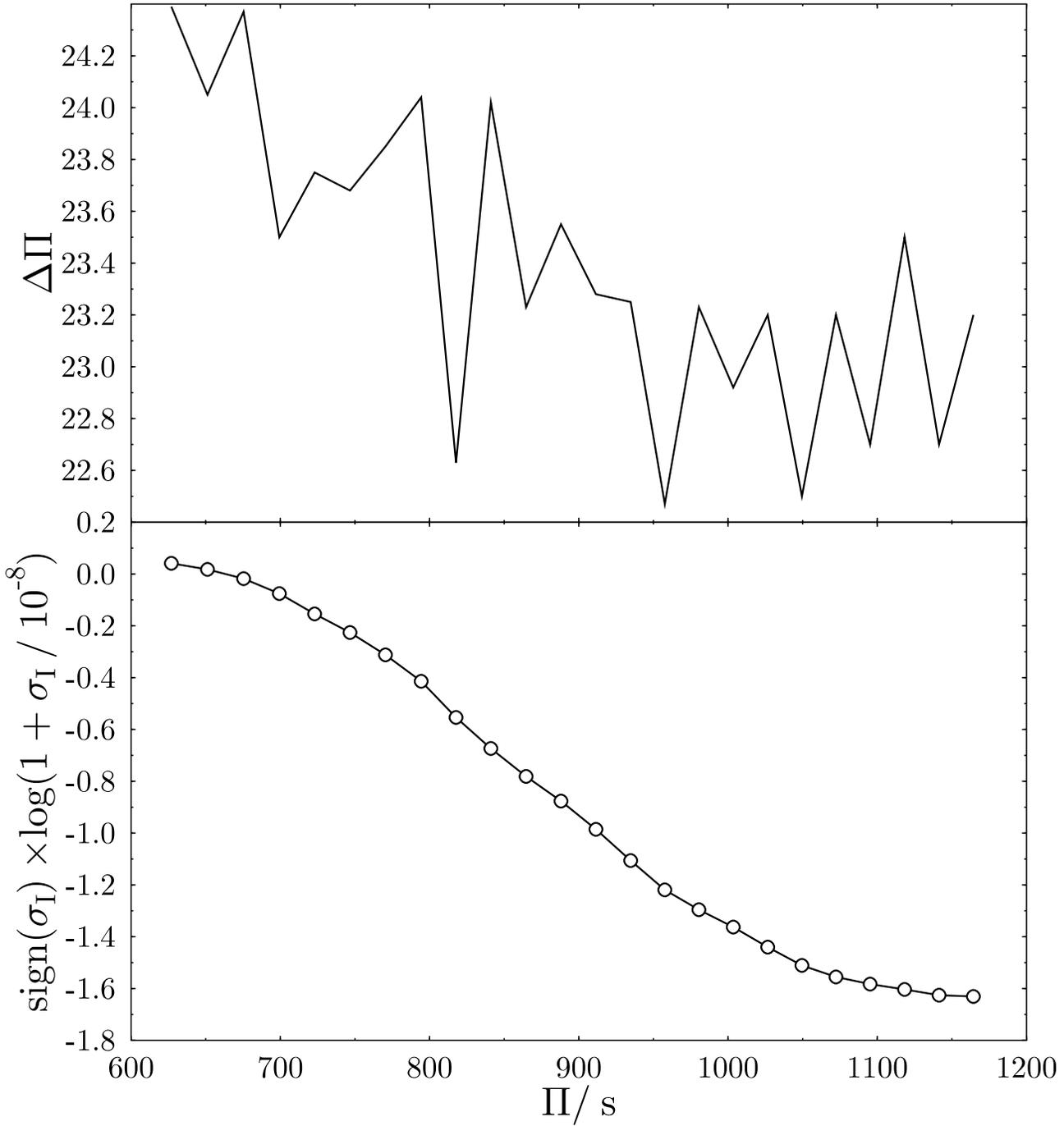
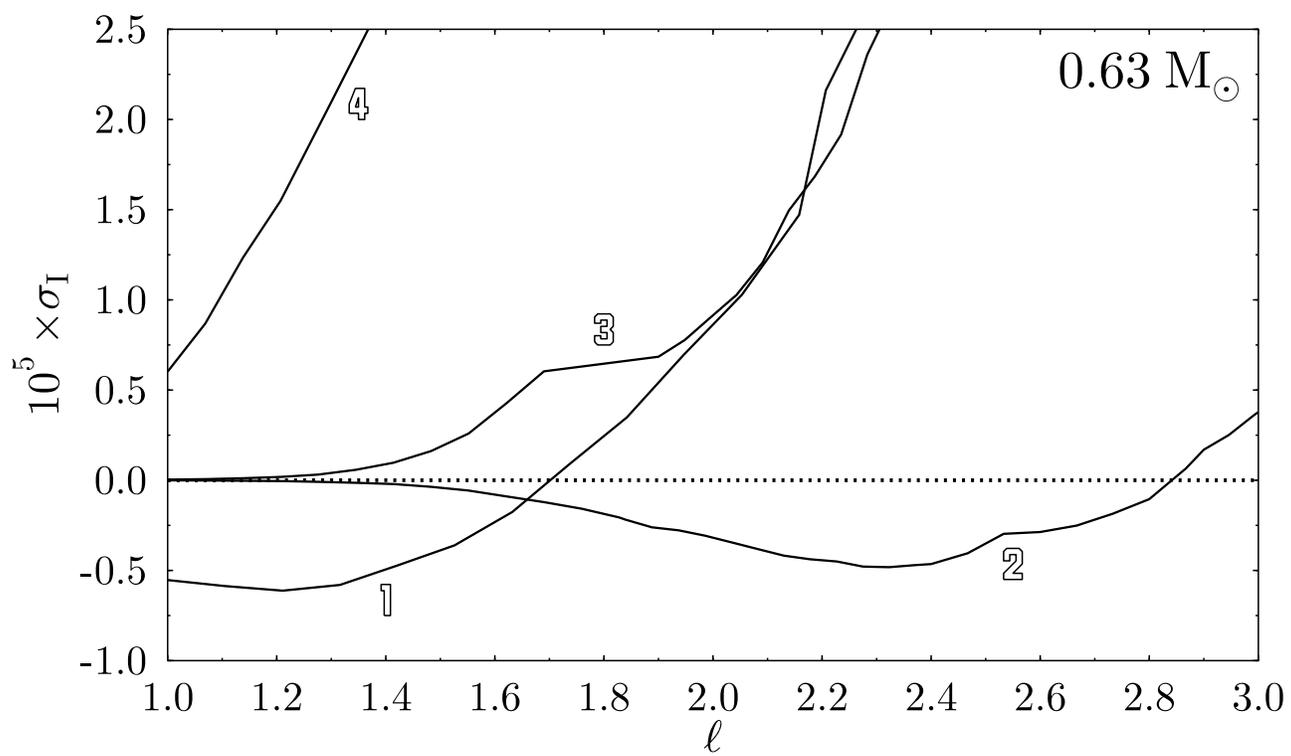
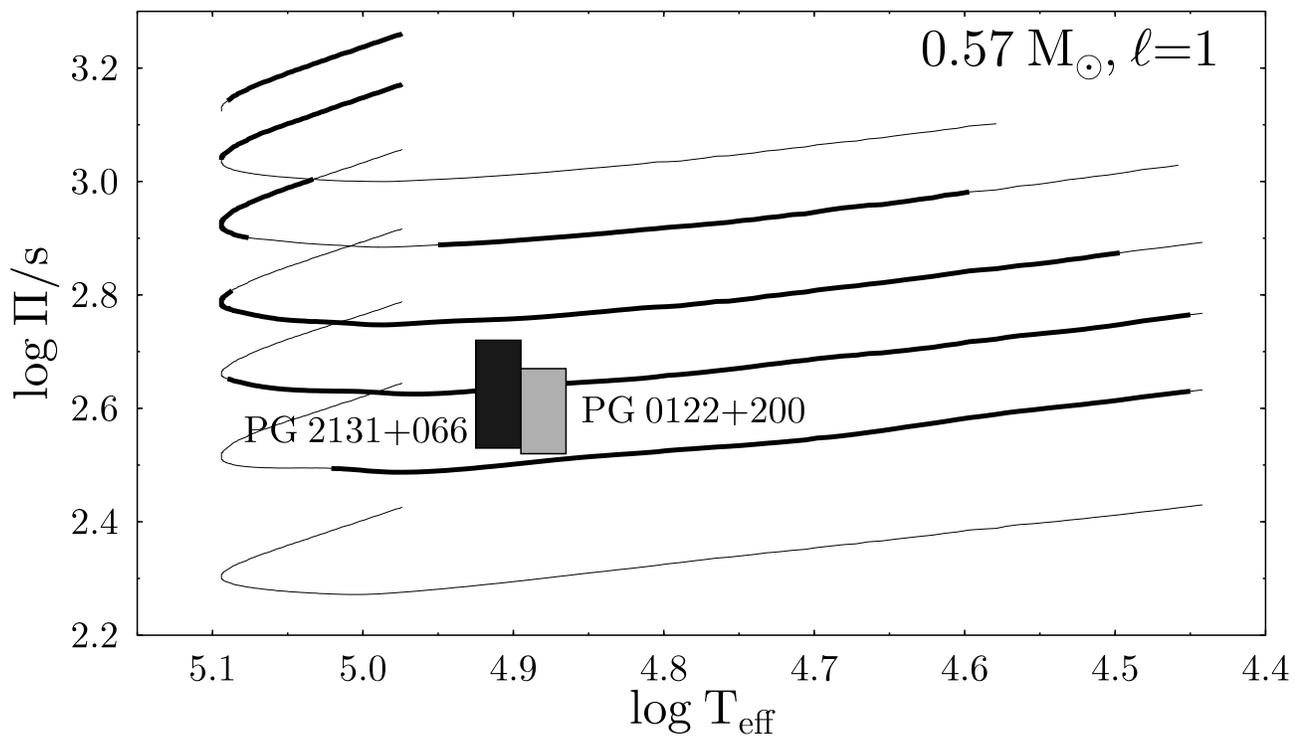
\bye